\documentclass[aps,pre,twocolumn,superscriptaddress,showpacs]{revtex4-1}
\usepackage{amssymb}
\usepackage{color}
\usepackage{graphicx}
\usepackage{amsmath}
\usepackage{mathrsfs}
\usepackage{times}
\usepackage{subeqnarray}
\usepackage{cases}
\usepackage{bm}
\usepackage[section]{placeins}
\usepackage{makecell}
\usepackage{appendix}

\usepackage[colorlinks,
            linkcolor=blue,
            anchorcolor=blue,
            citecolor=blue,
            urlcolor=blue,
            ]{hyperref}

\begin{document}

\title{Intercellular competitive growth dynamics with microenvironmental feedback}

\author{De-Ming Liu}
\affiliation{Institute of Computational Physics and Complex Systems, Lanzhou University, Lanzhou, Gansu 730000, China.}
\author{Zhi-Xi Wu}
\thanks{Corresponding author: wuzhx@lzu.edu.cn}
\affiliation{Institute of Computational Physics and Complex Systems, Lanzhou University, Lanzhou, Gansu 730000, China.}
\affiliation{Lanzhou Center for Theoretical Physics and Key Laboratory of Theoretical Physics of Gansu Province, Lanzhou University, Lanzhou, Gansu 730000, China}
\author{Jian-Yue Guan}
\affiliation{Institute of Computational Physics and Complex Systems, Lanzhou University, Lanzhou, Gansu 730000, China.}
\affiliation{Lanzhou Center for Theoretical Physics and Key Laboratory of Theoretical Physics of Gansu Province, Lanzhou University, Lanzhou, Gansu 730000, China}

\begin{abstract}

Normal life activities between cells rely crucially on the homeostasis of the cellular microenvironment, but aging and cancer will upset this balance.
In this paper, we introduce the microenvironmental feedback mechanism to the growth dynamics of multicellular organisms, which changes the cellular competitive ability, and thereby regulates the growth of multicellular organisms.
We show that the presence of microenvironmental feedback can effectively delay aging, but cancer cells may grow uncontrollably due to the emergence of the tumor microenvironment (TME).
We study the effect of the fraction of cancer cells relative to that of senescent cells on the feedback rate of the microenvironment on the lifespan of multicellular organisms, and find that the average lifespan shortened is close to the data for non-Hodgkin lymphoma in Canada from 1980 to 2015.
We also investigate how the competitive ability of cancer cells affects the lifespan of multicellular organisms, and reveal that there is an optimal value of the competitive ability of cancer cells allowing the organism to survive longest.
Interestingly, the proposed microenvironmental feedback mechanism can give rise to the phenomenon of Parrondo's paradox: when the competitive ability of cancer cells switches between a too high and a too low value, multicellular organisms are able to live longer than in each case individually.
Our results may provide helpful clues targeted therapies aimed at TME.

\end{abstract}
\date{\today }
\maketitle

\section{Introduction} \label{Introduction}

Competitive behavior is a universal phenomenon existing in all levels of biological population.
Animals compete with each other for territory, food, mates, and other resources~\cite{Wakefield2013,Harvey2013}.
The proliferation of root systems leads to competition within and between plants for nutrients and water in the surrounding environment~\cite{Adler2018}.
Competition for nutrients or oxygen also occurs among microbial populations~\cite{Baquero2003,rainey2003}. Generally, competitive behavior facilitates the selection of individuals better adapted to their environment, and enhances the evolutionary fitness of the population.
Specifically, competition exists between cells within multicellular organisms (especially cancer cells and their surrounding cells) by exerting changes in the microenvironment~\cite{Baker2020,flanagan2021,yum2021,Neerven2021}.
Intercellular competition is considered as an efficient mechanism to combat aging during biological evolution~\cite{Nagata2019,Amoyel2014}, where low competitive senescent cells are outcompeted by those higher competitive cells~\cite{Merino2015}.
Organisms can eliminate senescent cells through different apoptotic mechanisms (programmed cell death~\cite{Baker2017}, mainly induced by intercellular competition) to avoid the accumulation of harmful substances in senescent cells.

However, endless competition may lead to the overexploitation of environmental resources and thus to the tragedy of public (TOC)~\cite{Hardin1968}.
Examples include uncontrolled population growth~\cite{Hardin1968}, climate problems~\cite{Zhang2017}, groundwater depletion due to excessive irrigation~\cite{Aeschbach2012}, etc.
In general, TOC causes the system to enter a degraded state of shared resources, which adversely affects the evolution and ecological health of all individuals in the long term.
TOC also occur likely in intercellular competition, where those cells with high competitive ability may attack functional cells, reflecting the \emph{ambition} to reproduce indefinitely, which is usually considered as a mechanism for the generation of cancer cells~\cite{Capp2021}.
Taken together, organisms effectively remove part of the senescent cells through intercellular competition and slow down the rate of loss of functional cells, thus extending lifespan, but with the risk of accelerating the production of cancer cells, making the aging of multicellular organisms inevitable~\cite{Baillon2014, Nelson2017}.

We notice that real-world complex systems are typically characterized by bidirectional feedback between the environment and the strategy interaction incentives~\cite{Tilman2020}.
For example, the temperature of the environment~\cite{Reinke2022} and the status of environmental resources~\cite{Silva2022} are key factors that affect the lifespan of the organisms.
Changing strategy behavior in terms of the environmental feedback can effectively avoid TOC~\cite{Weitz2016}.
Thus, environmental feedback is an essential mechanism to resovle TOC.
Motivated by this point, in this work we ask whether there are similar environmental feedback mechanism in multicellular organisms that would balance aging and cancer to keep the organism alive for longer?

Homeostasis in multicellular organisms (that is, maintenance of the steady state that is essential for the survival of the organism) depends on a high degree of intercellular cooperation~\cite{Wen2022PRL}.
For each cell, dynamic equilibrium is widely present in the microenvironment in which it lives~\cite{Meng2021}.
Organisms provide energy for cellular function and growth by consuming mircoenvironmental resources for metabolism~\cite{Heiden2009}.
Energy for the preservation of normal cellular functions, such as growth, survival, apoptosis, differentiation and morphogenesis, is provided by the microenvironment~\cite{Liu2010PNAS,Vogel2006}. However, aging and cancer are accompanied by changes in the state of the microenvironment.
For instance, organisms suffering from metabolic diseases arising from aging can lead to microenvironmental imbalances~\cite{Wiley2021}.
Cancer cells have higher levels of glycolysis than normal cells~\cite{Park2020}, and rapid reproduction of cancer cells thus consumes a lot of energy~\cite{Papalazarou2018}.
Immune cells and cancer cells compete intensely for nutrients in TME, which is unsuitable for the survival of most functional cells~\cite{Anderson2020,Reinfeld2021}.

Based on the biological facts that TME is not suitable for functional cell growth~\cite{Anderson2020,Reinfeld2021} and intercellular competition can remove senescent cells~\cite{Nelson2017}, we propose intercellular competitive growth dynamics with microenvironmental feedback.
The work of Wen {\it{et al.}} manifested that cells in a state of alternating competitive and cooperative switching (by taking into account the effects due to periodic rhythms) can extend lifespan~\cite{Wen2022PRL}.
Their results inspired us to recognize that the growth dynamics of cells may be related to the environmental state they are living in.
For instance, tumour progression is profoundly influenced by the interactions of cancer cells with their environment~\cite{Truffi2020}.
Therefore, we let the microenvironmental state react on the competitive ability of functional cells according to the characteristics of functional cells in different microenvironmental states.

This paper is organized as follows.
In Sec.~\ref{model}, we introduce cellular competitive ability, microenvironmental feedback mechanisms, dynamics and production of all types of cells in our model.
Numerical results and analysis are given in Sec.~\ref{numerical results and discussions}, where we explain in detail the effect of the microenvironmental feedback parameters on the lifespan.
Finally, we briefly summarize this work and give some relevant discussions in Sec.~\ref{conclusions}.

\section{model} \label{model}

\begin{figure}[htb]
\begin{center}
\includegraphics[width=0.4\textwidth]{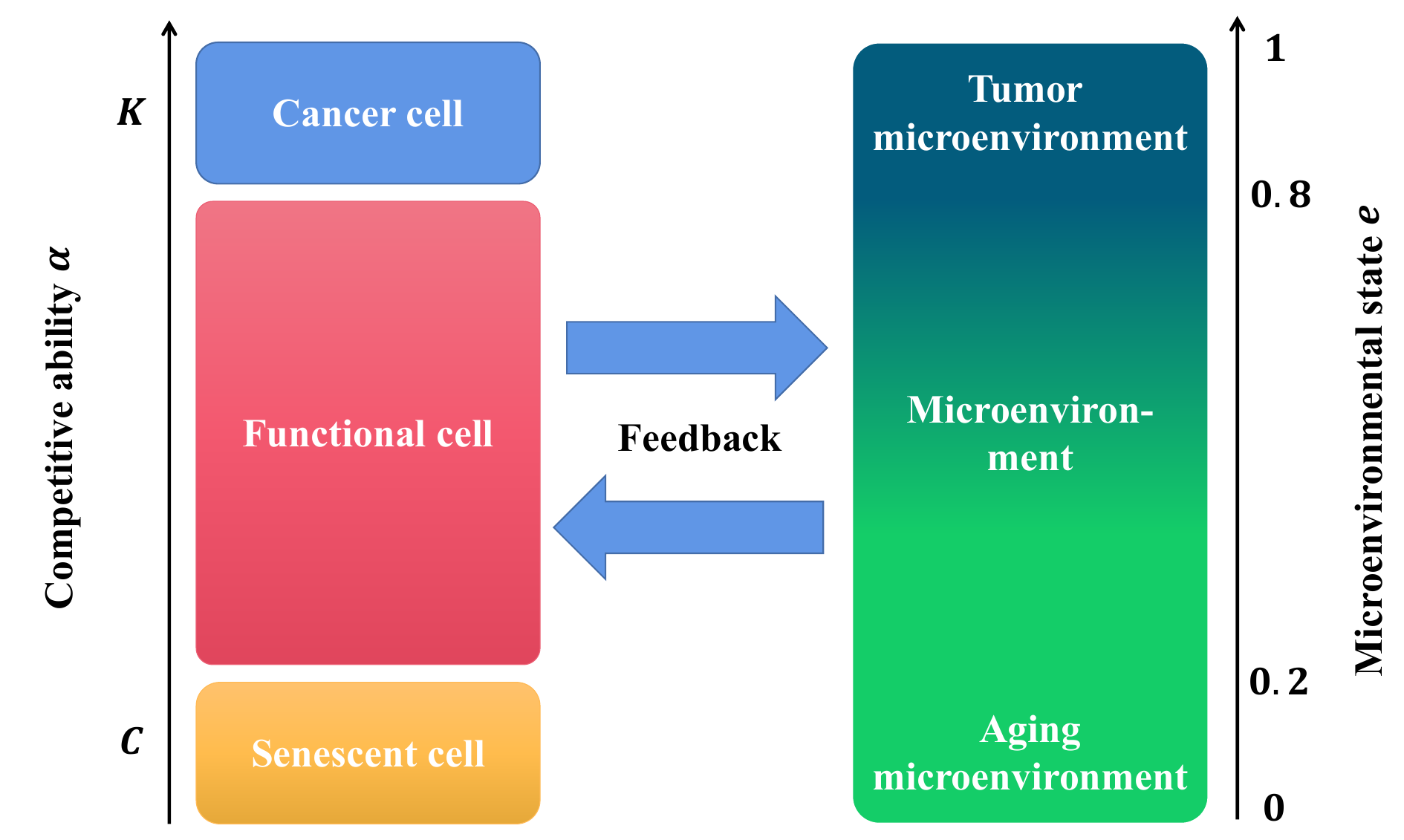}
\caption{
Schematic diagram of the microenvironmental feedback. 
On the left are three types of cells divided according to their competitive ability, where the competitive ability of the functional cells is regulated by the state of the microenvironment. 
On the right are the senescent, normal and tumour microenvironments, which are distinguished by the microenvironmental state parameter $e$. 
The fraction of cells in the multicellular organism can influence the state of the microenvironment.
}
\label{fig1}
\end{center}
\end{figure}

In order to describe the relationship between cell growth and microenvironment in multicellular organisms, we extend the model proposed in Ref.~\cite{Nelson2017} by including a microenvironmental feedback mechanism.
As shown in Fig.~\ref{fig1}, the model consists of two parts: multicellular growth dynamics and evolution of the state of the microenvironment.
The composition of various types of cells in organisms constitutes the state of the microenvironment (being as aging, normal or TME), which in turn regulates the growth rate of functional cells.

\subsection{Competitive ability of each cell type}

To investigate the intercellular competition, we classify the cells in an organism into three types as in previous work~\cite{Nelson2017}, namely, cancer cells, functional cells and senescent cells, whose competitive ability $\alpha$ is assumed to be from high to low: $\alpha_{C}>\alpha_{F}>\alpha_{S}$.
For simplicity, we suppose that the competitive abilities of senescent and cancer cells are totally determined by the characteristics of the cells themselves, and microenvironment changes can hardly affect their competitive ability at the cellular level due to the low activity of senescent cells and the infinite reproduction capacity of cancer cells.
Therefore, their competitive abilities are assumed to be independent of the microenvironmental state.
Without loss of generality, we presume that senescent cells have low competitive ability ($\alpha_S=C$) and cancer cells have high competitive ability ($\alpha_C=K$), where $C\ll K$.
By contrast, functional cells are presumed to be sensitive to the changes in the state of the microenvironment.
Following common practice~\cite{Weitz2016}, we define the competitive ability of functional cells as:
\begin{eqnarray}
 \alpha_F=e(C+\delta)+(1-e)(K-\delta),
 \label{1}
\end{eqnarray}
where $e \in[0,1]$ measures the state of the microenvironment.
The greater $e$ is, the more the microenvironmental state tends toward to TME.
Conversely, the lower the $e$ is, the more the microenvironmental state tends toward to the aging microenvironment.
The parameter $\delta$ is introduced to describe the competitive advantage of functional cells versus senescent cells, or cancer cells versus functional cells in extreme conditions.
For example, the competitive ability of functional cells is $\alpha_F=C+\delta>\alpha_S$ and $\alpha_F=K-\delta<\alpha_C$ when $e=1$ and $e=0$, respectively.
Therefore, the values of the parameter $\delta$ must satisfy $C+\delta<K-\delta$.
To conveniently depict the specific states of microenvironment at different stages, we choose $e\in \left\lfloor0.8, 1\right\rfloor$ as TME and $e\in \left\lfloor0, 0.2\right\rfloor$ as the aging microenvironment.
We have checked that other choices of the intervals of $e$ do not change the qualitative properties of the results presented below.

\subsection{Changes in microenvironmental state}

We denote the fraction of each cell type in the organism by
\begin{eqnarray}
f_{a} & = & \frac{n_{a}}{n_{F}+n_{S}+n_{C}}, \quad a \in\{F, S, C\},
\label{2}
\end{eqnarray}
where $n_F$, $n_S$, and $n_C$ are the number of functional, senescent, and cancer cells, respectively.
Then, the change in microenvironmental state is determined by the variation of the fraction of each cell type.
Functional cells ensure the microenvironment to be in homeostasis because the normal life activity of the organism cannot be separated from the dynamic balance of matter and energy within the microenvironment~\cite{Huang2017}.
However, aging and cancer are often accompanied by changes in the microenvironment.
For instance, cancer cells get TME suitable for survival (for uncontrolled growth of cancer cells)~\cite{Hannafon2013, Arneth2020}.

Based on the above biological facts, we argue that when the fraction of senescent cells in the organism increases, it will degrade the normal microenvironment and lead to the emergence of aging microenvironment.
Similarly, when the fraction of cancer cells in the organisms increases, it deteriorates the normal microenvironment resulting in the arise of TME.
By taking these two factors into account, we assume that the microenvironmental state $e$ is modified by composition of the cellular population:
\begin{eqnarray}
e(t+1) & = & e(t)+\epsilon  e(t)[1-e(t)]\left( \theta f_{C}-f_{S} \right),
\label{3}
\end{eqnarray}
where $\theta$ indicates the relative ratio of the feedback strength of cancer cells and senescent cells on the microenvironment.
With large $\theta$, one expects that the organism is easily to evolve to the TME state in the presence of cancer cells.
As in~\cite{ Weitz2016,Bairagya2021}, the dimensionless $\epsilon$ quantity satisifies $\epsilon \ll 1$ such that the changes in the microenvironment at the organismal level are much slower as compared to those in the cellular fraction of the multicellular organism.

\subsection{Growth dynamics with intercellular competition }

The changes in the number of cells $n_a$ within one developmental time step is described by a growth model with intercellular competition~\cite{Nelson2017}:
\begin{eqnarray}
n_{a}(t+1)=r\left[p \frac{\alpha_{a}}{\bar{\alpha}}+(1-p)\right] n_{a}(t),
\label{4}
\end{eqnarray}
where $\bar{\alpha}=\sum_{a} \alpha_{a} n_{a} / \sum_{a} n_{a}$, $r$ is the natural growth rate of the cell, and $r > 1$ stands for the intrinsic growth rate.
When all types of cells grow in the absence of competition ($p = 0$), multicellular organisms exhibit cooperative growth.
For $p \neq 0$, multicellular organisms exhibit competitive growth because $\alpha_a$ depends on the type of cells, refer to Eq.~(\ref{1}).

In the initial stage, all cells are functional and there are no senescent or cancer cells ($n_{S}(0)=n_{C}(0)=0$).
As growth and development progress, part of the functional cells inevitably get aging.
Besides, external stimuli and/or genetic mutations may lead to the arise of cancer cells.
Accounting for these two points, we describe the dynamics of growth of each cell type with microenvironmental feedback by the following set of equations:
\begin{eqnarray}
\begin{aligned}
n_{F}(t+1)= & r\left[p \frac{\alpha_{F}}{\bar{\alpha}} +(1-p) \right]n_{F}(t) \\&-\mu_{F S} n_{F}(t)-\mu_{F C} n_{F}(t), \\
n_{S}(t+1)= & r\left[p \frac{\alpha_{S}}{\bar{\alpha}} +(1-p) \right]n_{S}(t)+\mu_{F S} n_{F}(t), \\
n_{C}(t+1)= & r\left[p \frac{\alpha_{C}}{\bar{\alpha}} +(1-p) \right]n_{C}(t)+\mu_{F C} n_{F}(t).
\end{aligned}
\label{5}
\end{eqnarray}
Here, $\mu_{FS}$ is the probability that the functional cells are changed into senescent cells (including inevitable aging and mutation).
$\mu_{FC}$ is the probability that functional cells mutate to cancer cells.

Equations~(\ref{3}-\ref{5}), together with Eq.~(\ref{1}) as the link between microenvironmental state and the competitive ability of the functional cells, constitute the set of dynamical equations that mathematically describes intercellular competitive growth dynamics with microenvironmental feedback. Solving these equations numerically, we are able to obtain how the number of each cell type evolve with developmental time step.

\section{Results and analysis} \label{numerical results and discussions}

Aging is a process in which the function of various tissues and organs of the organism decrease comprehensively with the increase of age, which is shown to be inevitable due to the intercellular competition~\cite{Nelson2017}.
The main problem we are interested in is the effect of intercellular competition and microenvironmental feedback on the lifespan of multicellular organisms.
To this end, we define the time at which the fraction of functional cells decreases to $\lambda=1\%$ as the lifespan $L$ of organisms, as done in~\cite{Wen2022PRL}.
We calculate the fitness (or competitive ability) of each cell type under different conditions, where the parameters are chosen as $C=1$, $K=10$, $\delta=1$, $r=1.001$, $\epsilon=0.01$, $\mu_{FS}=2 \times 10^{-3}$, and $\mu_{FC}=5 \times 10^{-5}$ if not otherwise specified.
In the initial stage of development, all cells are functional so that the initial number of each cell type is $n_{F}(0)=10$, $n_{S}(0)=n_{C}(0)=0$.
The results presented below are obtained for the initial state of the microenvironment with $e(0)=0.5$.
In the Appendix~\ref{Impact of lambda and initial value of microenvironment state on lifespan.}, we also show the impact of $\lambda$ and other initial values of the microenvironment state on the lifespan.

\subsection{Intercellular competition can combat aging without microenvironmental feedback} \label{Intercellular competition can combat aging}

\begin{figure*}[htb]
\begin{center}
\includegraphics[width=0.30\textwidth]{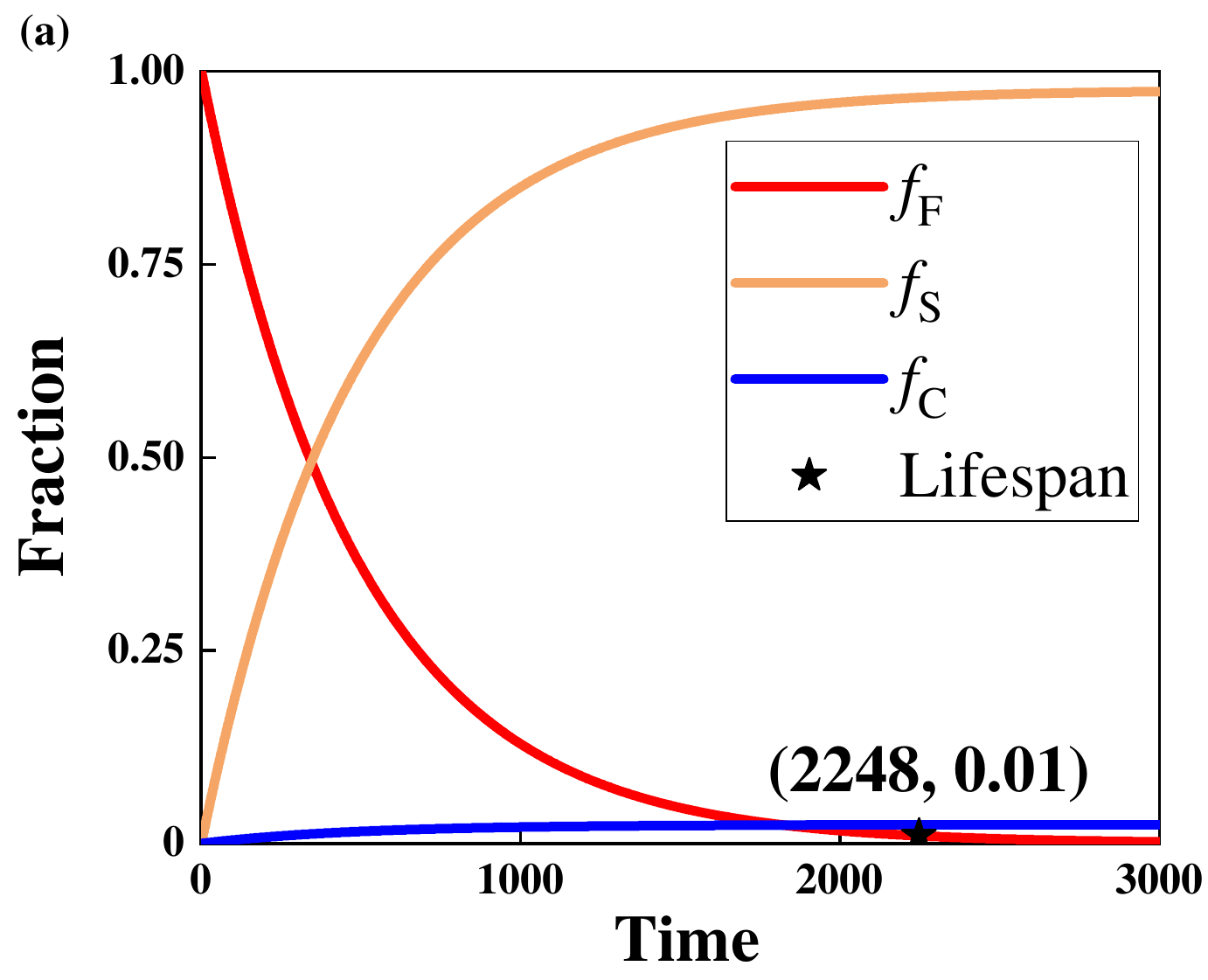}
\includegraphics[width=0.30\textwidth]{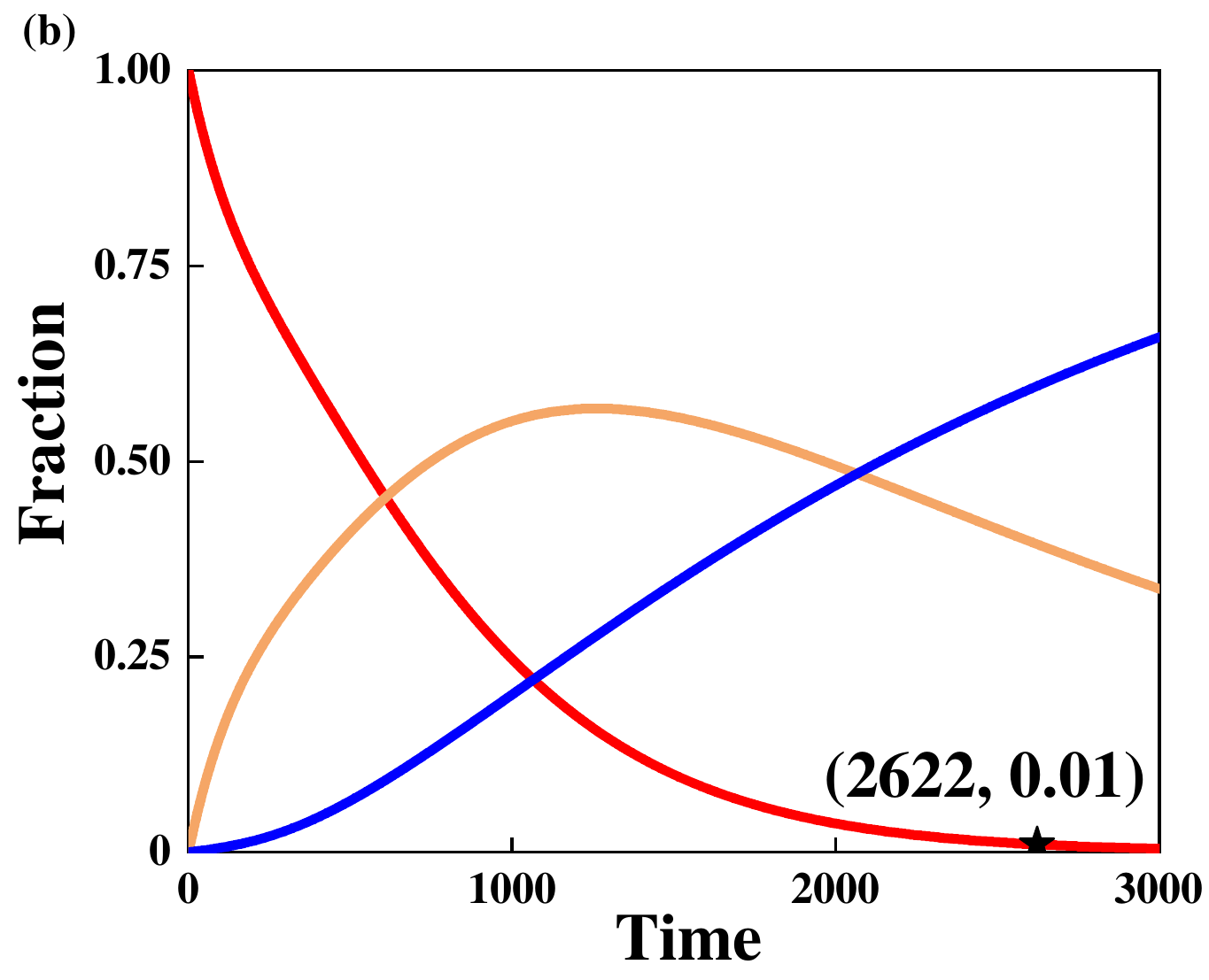}
\includegraphics[width=0.30\textwidth]{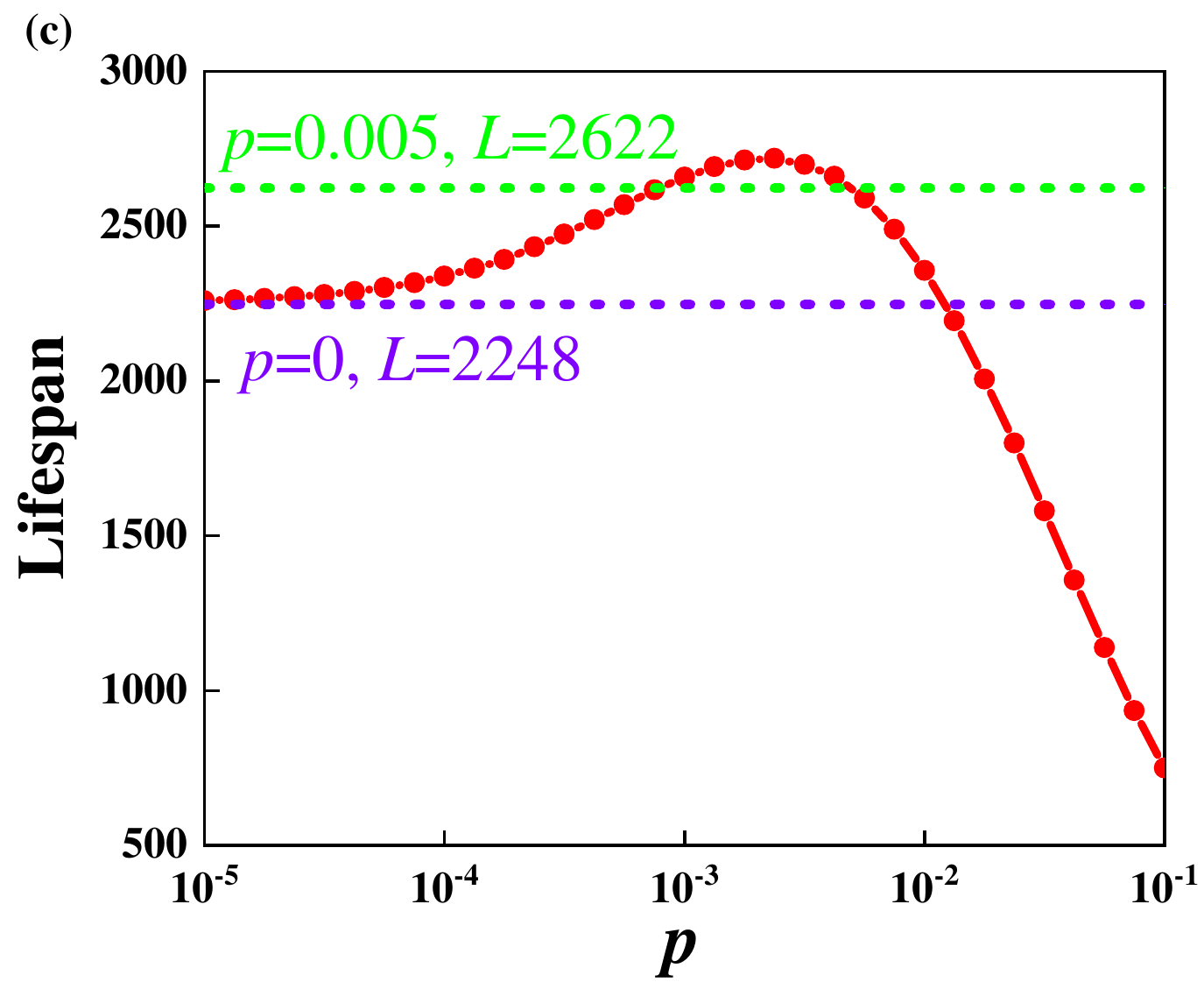}
\caption{
Fraction of different types of cells as a function of time, where (a) without intercellular competition ($p=0$), all cells grow under natural growth rate condition; and (b) with intercellular competition ($p=0.005$), intercellular competition successfully delays the aging of the organism ($L= 2622$).
(c) The lifespan $L$ of the multicellular organism as a function of the competition parameter $p$.}
\label{fig2}
\end{center}
\end{figure*}

First, we present in Fig.~\ref{fig2} the effects of intercellular competition on the lifespan of the multicellular organism.
Without intercellular competition ($p=0$), the fraction of functional cells decreases to $1\%$ at $t=2248$, and senescent cells dominate in the organism as shown in Fig.~\ref{fig2}(a).
The increase in the fraction of senescent cells is inhibited by the growth of cancer cells when partial competitive growth ($p=0.005$) is introduced in Fig.~\ref{fig2}(b).
The lifespan of the multicellular organism is largely extended as compared to the situation without intercellular competition.
This suggests that intercellular competition can combat aging, but the decrease in the fraction of functional cells is not altered, since those high competitive cancer cells have evident growth advantage on both low competitive senescent cells and functional cells.

However, excessive competition among the cells is found to be detrimental for the multicellular organism.
Figure~\ref{fig2}(c) displays how the lifespan of the multicellular organism changes as the proportion of competition increases.
When the percentage of competition $p$ is too high, the lifespan of the multicellular organism is shortened compared to the case of without intercellular competition.
Intercellular competition likely leads to the dominance of cancer cells in the organism, and the development of too many cancer cells gives rise to a shortened lifespan of the organism.
This result is in accordance with the findings in~\cite{Nelson2017}, where intercellular competition creates an inescapable ``double bind" that makes aging inevitable in multicellular organisms.

\subsection{Microenvironmental feedback can extend lifespan} \label{Microenvironmental feedback can extend lifespan.}

\begin{figure*}[htb]
\begin{center}
\includegraphics[width=0.30\textwidth]{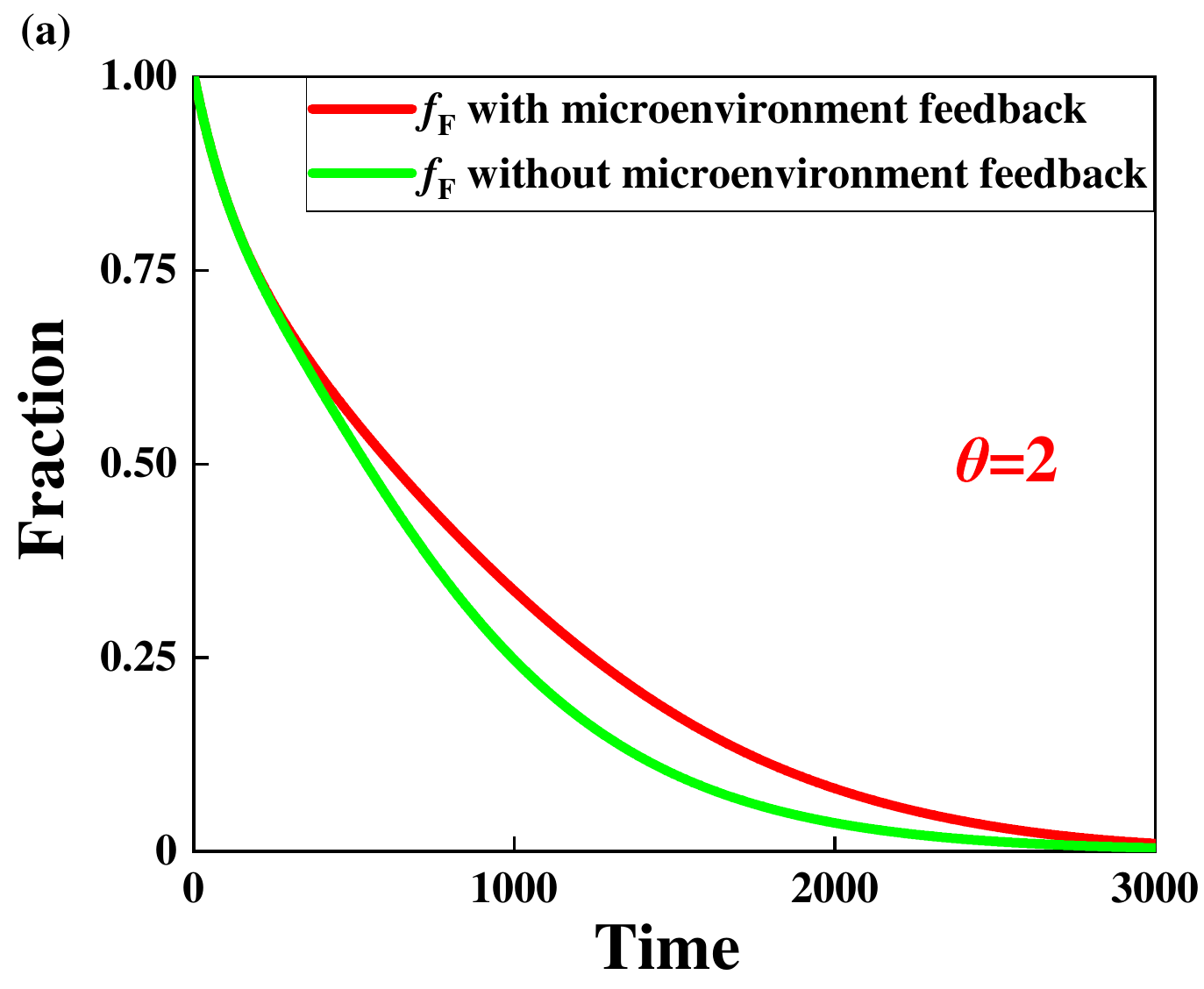}
\includegraphics[width=0.30\textwidth]{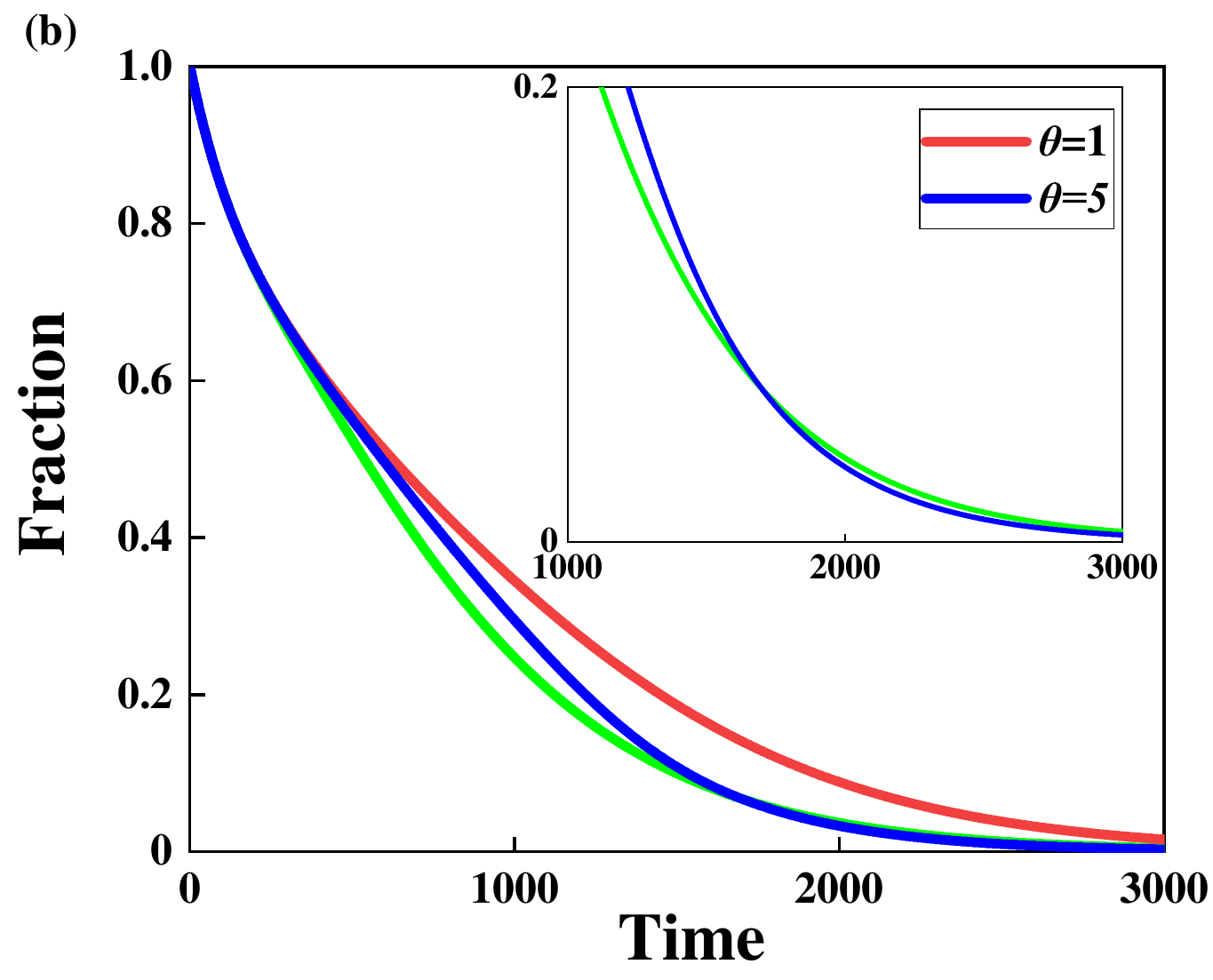}
\includegraphics[width=0.30\textwidth]{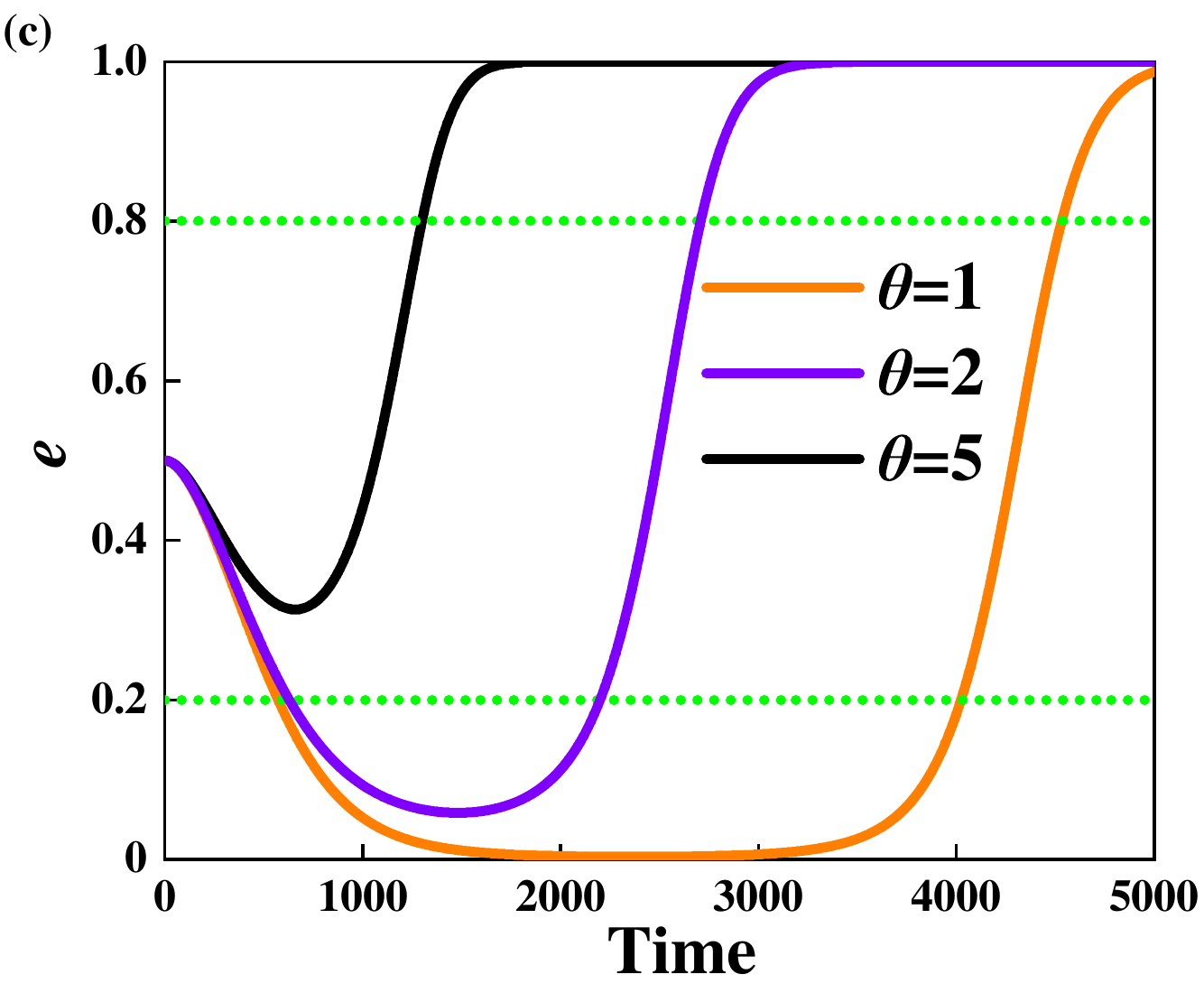}
\caption{
Fraction of the functional cell as a function of time, where (a) in the presence of intercellular competition with microenvironment feedback ($\theta=2$), we have $L_{\theta=2}=3005$; and (b) under the cases of $\theta=1$ and $\theta=5$, we have $L_{\theta=1}=3236$, $L_{\theta=5}=2493$, respectively.
(c) Time evolution of the microenvironmental state $e(t)$ for the cases of $\theta=1$, $2$, and $5$.}
\label{fig3}
\end{center}
\end{figure*}

We now discuss how the introduction of microenvironmental feedback affects the lifespan of the organism.
To do this, we fix the parameter $p = 0.005$ and check the dynamic behavior of the fraction of different types of cells as well as that of the microenvironment state for different choices of $\theta$.
Figure~\ref{fig3}~(a) features the time series of the fraction of functional cells for $\theta=2$.
We note that the lifespan is successfully extended due to the presence of microenvironmental feedback.
As expected from our model, when the fraction of senescent cells in the organism increases, functional cells are affected by the changes in the microenvironmental state to increase their competitive ability to better delay aging.
However, we find that for sufficiently large $\theta=5$, the microenvironmental feedback accelerates aging instead, as shown in Fig.~\ref{fig3}(b).
This suggests that the existence of microenvironmental feedback has a non-trivial effect on the lifespan.
We present in Fig.~\ref{fig3}(c) the evolution of the microenvironmental state for three values of $\theta=1$, $2$, and $5$, respectively.
We find that larger $\theta$ usually pleads to faster evolution to TME of the organism.
Notice that the competitive ability of functional cells is significantly reduced in TME.
Thus, although senescent cells are efficiently removed due to the presence of microenvironmental feedback, cancer cells and TME dominate the multicellular organism leading to the inevitable reduction in the fraction of functional cells.

By checking the evolution of the microenvironmental state, we can better understand how microenvironmental feedback extends lifespan.
When intercellular competition exists (see in Fig.~\ref{fig2}(b)), the fraction of senescent cells in the organisms is highest among the three types of cells at around $t\in \left\lfloor600,2000\right\rfloor$, and eventually the fraction of cancer cells will dominate the organism.
Consequently, the microenvironmental state will change from normal to aging microenvironment, and finally to the TME.
After introducing the microenvironmental feedback, we find that the case of $\theta = 2$ is in the best accordance with the above microenvironmental evolutionary rule.

The rate of decrease in the fraction of functional cells is almost the same under different conditions at $t<300$, see Fig.~\ref{fig3}(a).
The microenvironmental changes are smaller at this period, so it is difficult for the environmental feedback mechanisms to play a role on the aging process of multicellular organisms.
When the microenvironmental state changes to aging microenvironment at $t \in  \left\lfloor629,2200\right\rfloor$, the competitive ability of the functional cells is increased substantially and aging is delayed.
As the fraction of cancer cells increases, the microenvironment eventually evolves to the TME ($t>2708$) suitable for infinite growth of cancer cells.

\begin{figure}[htb]
\begin{center}
\includegraphics[width=0.4\textwidth]{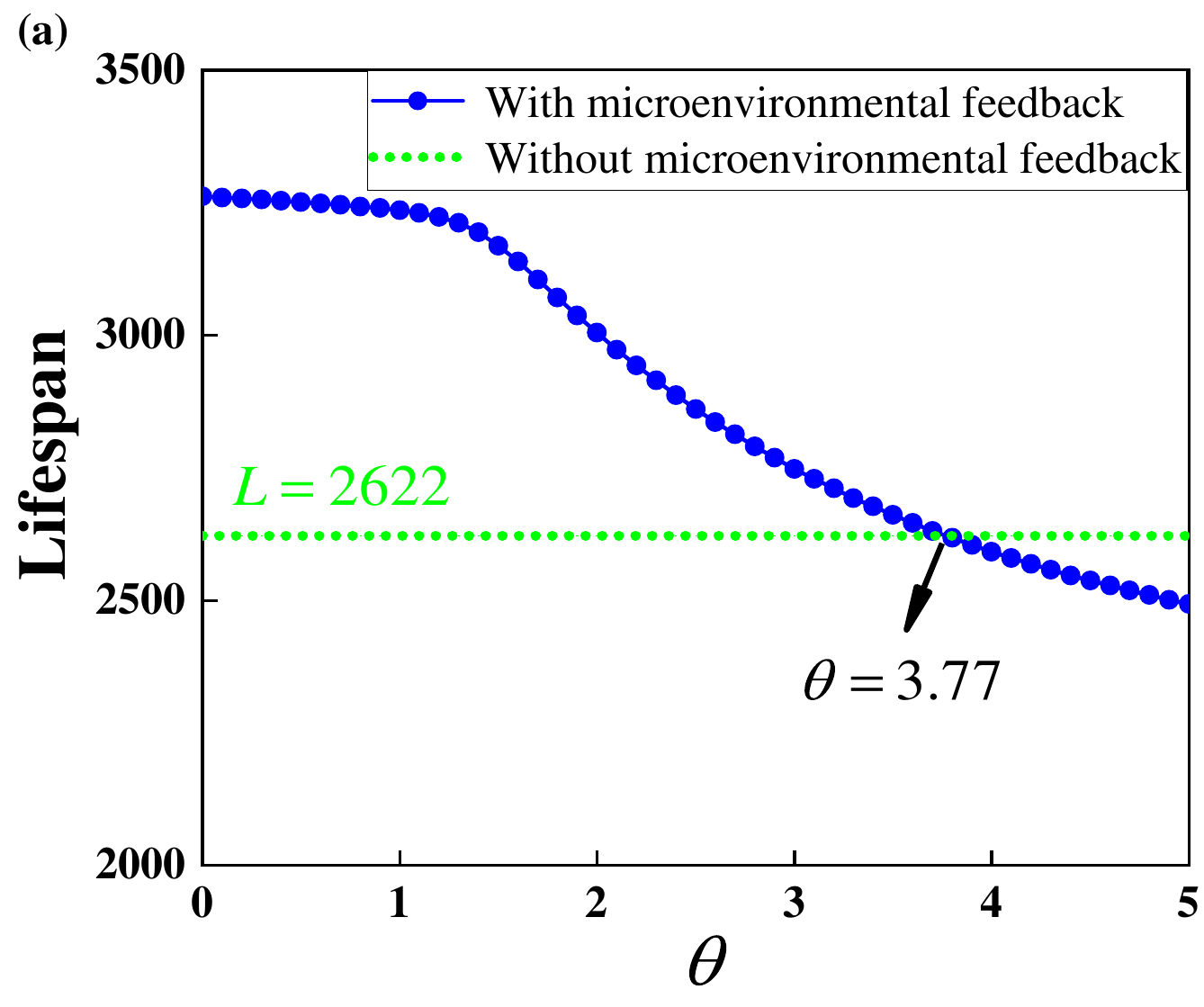}
\includegraphics[width=0.4\textwidth]{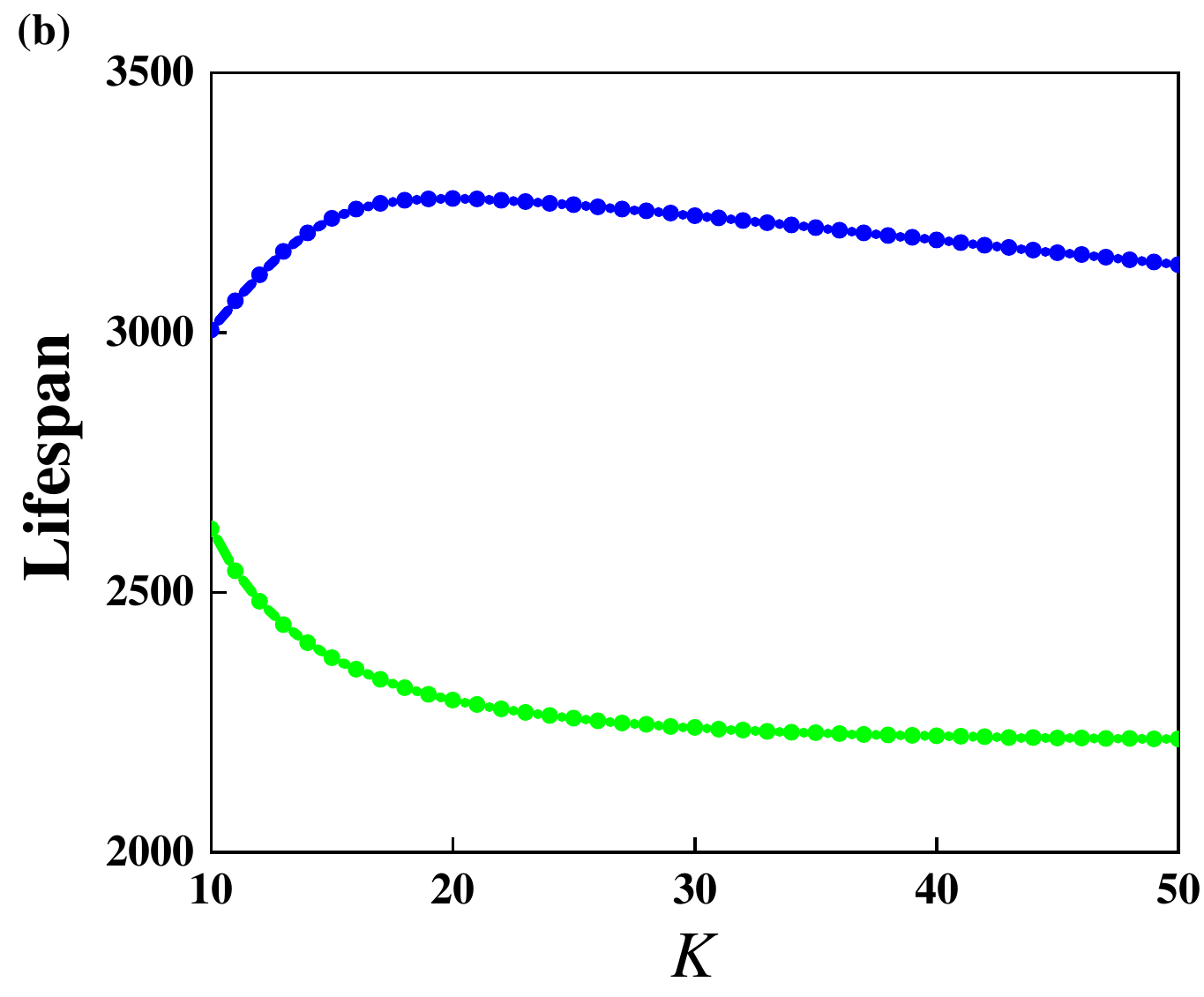}
\caption{
Lifespan $L$ under different parameters $\theta$ and $K$.
(a) Lifespan $L$ as a function of the feedback strength $\theta$ of the cancer cells' fraction to the microenvironment, with $K=10$.
(b) Lifespan $L$ as a function of the competitive ability $K$ of the cancer cells, with $\theta=2$.}
\label{fig4}
\end{center}
\end{figure}


\subsection{Effect of microenvironmental feedback parameters on lifespan}
\label{Effect of microenvironmental feedback parameters on lifespan.}

We then study the influence of the feedback rate $\theta$ and the competitive ability of cancer cells $K$ on the lifespan $L$ of the multicellular organisms.
The results summarized in Fig.~\ref{fig4}(a) show that the lifespan $L$ decreases monotonically with the parameter $\theta$ due to the enhanced feedback of cancer cells' fraction to the microenvironment (faster speed to reach to the state of TME for larger $\theta$).
Using the case without microenvironmental feedback as a baseline, we observe a threshold phenomenon that the microenvironmental feedback can effectively extend the lifespan for $\theta<3.77$.
We notice that the microenvironmental feedback mechanism has little effect on the lifespan for $\theta<1$, as cancer is less influential than aging in such situation.
When $\theta$ is between $1$ and $3.77$, we see that the microenvironmental feedback can extend properly the lifespan as compared to the case of without microenvironmental feedback.
However, too strong feedback rate of the fraction of cancer cells relative to the fraction of senescent cells to the microenvironment ($\theta>3.77$) leads to the rapid development of the TME triggering malignant or acute tumors, which causes shortened lifespan instead.

It is worth pointing out that the parameter $\theta$ is a key measure of how quickly TME emerges.
In recent years, targeted therapies to address TME have made rapid progress~\cite{Chen2015,Quail2017,Jiang2020,Xiao2021}.
In this regard, we consider $\theta$ to be a significant candidate parameter for measuring the effectiveness of treatments targeting TME.
The shortened lifespan is a highly visual statistic in cancer treatment. 
Pham and coworkers found that men with non-Hodgkin lymphoma lost an average of $23.7\%$ of their lifespan in 1980 versus $16.1\%$ in 2015, while women with non-Hodgkin lymphoma lost an average of 21.7\% of their lifespan in 1980 versus $15.5\%$ in 2015~\cite{Pham2020}.
From 1980 to 2015, improvements in medical care can be regarded as a key factor in the reduction in the shortened lifespan of cancer patients.
In our model, $\theta$ is the parameter that reflects the rapidity of TME emergence, so that delaying the emergence of TME (reducing $\theta$) could be considered as a targeted treatment for TME. 
It is reasonable for us to argue that the absence (or the delay of emergence) of TME is the most optimal effect of therapies targeting TME.
Corresponding to the case of $\theta = 0$ in our model: $L_{\theta=0}=3262$. 
The shortened lifespan $\kappa$ can be quantitatively described as:
\begin{eqnarray}
\kappa_\theta  =\frac{L_0 -L_\theta}{L_0} \times 100\%.
\label{6}
\end{eqnarray}
We notice that the results for $\theta=3$ and $\theta=5$ on the two sides of the threshold ($\theta=3.77$) are close to the data in the work of Pham \emph{et al.}~\cite{Pham2020}.
Therefore, the parameter $\theta$ can measure the change in these two time points at the macro level.  This suggests that our model may serve as a useful evaluation system for therapeutic approaches targeting TME.

\begin{figure}[htb]
\begin{center}
\includegraphics[width=0.4\textwidth]{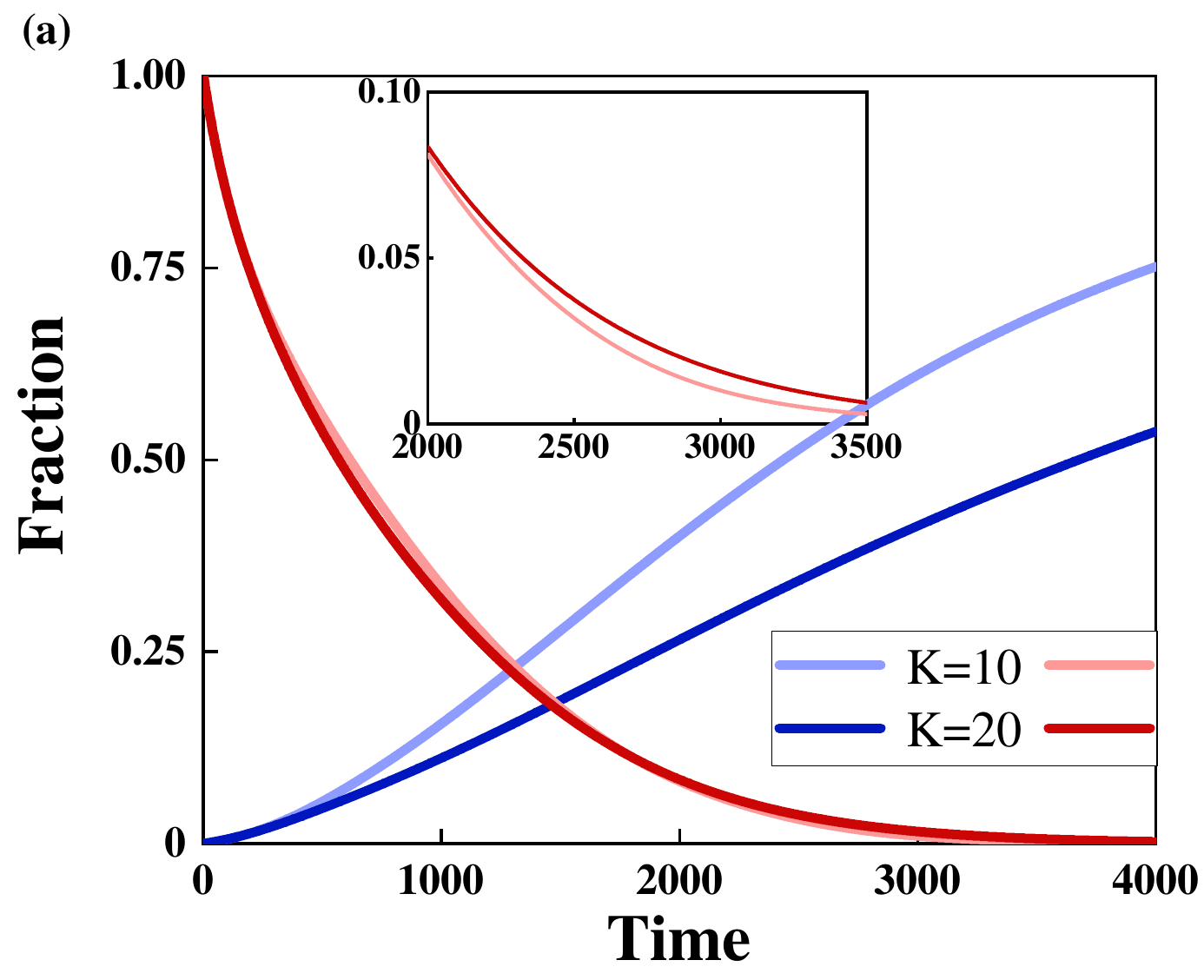}
\includegraphics[width=0.4\textwidth]{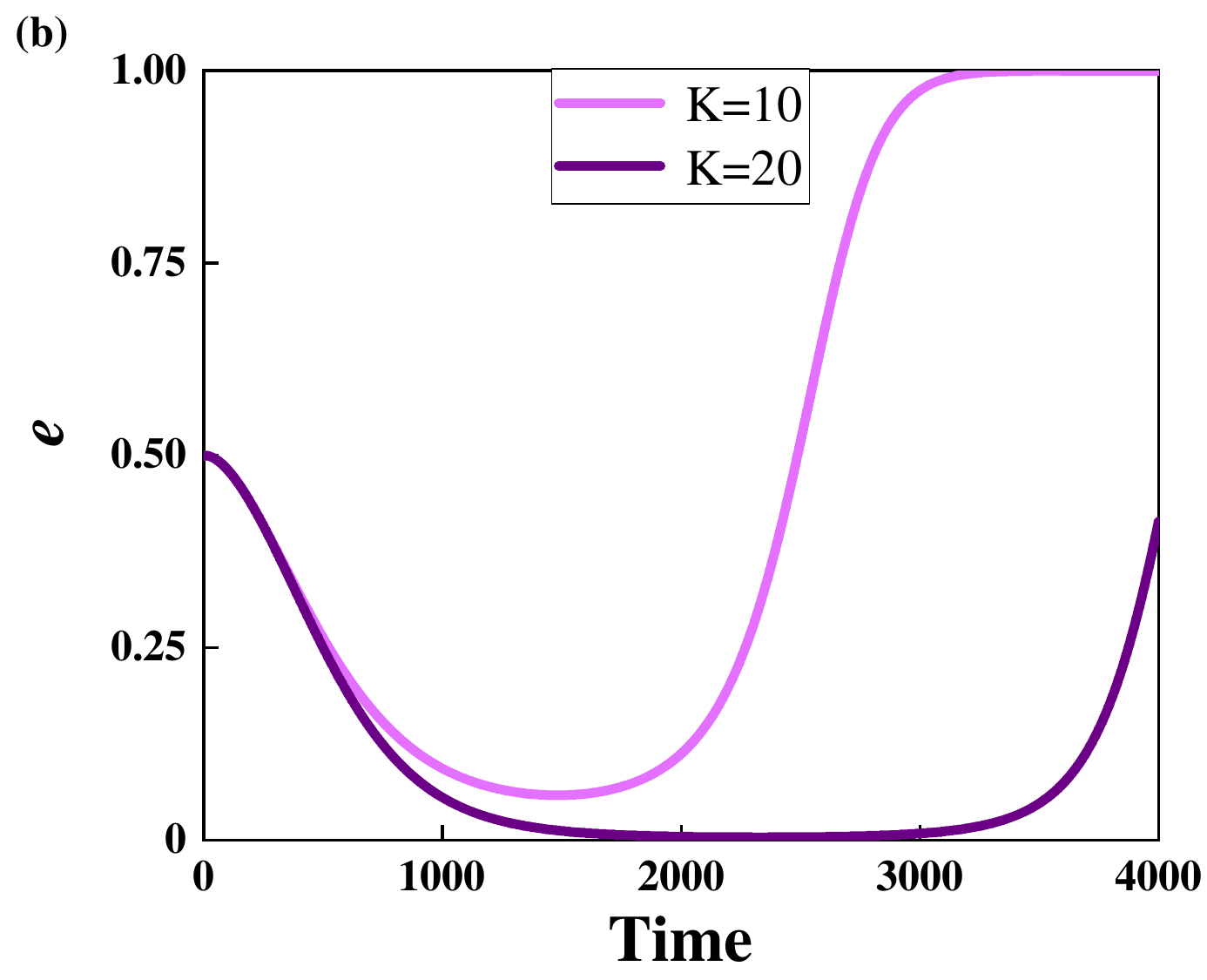}
\caption{
Time series of the fraction of the functional cells (red line) and the cancer cells (blue line) for two different choices of $K$, with $\theta=2$.
The relative magnitude of the reduction rate of the functional cells for $K=10$ and $K=20$ changes on both sides of $t=1795$.
(b) Time evolution of the microenvironment state under $K=10$ and $K=20$.}
\label{fig5}
\end{center}
\end{figure}

We next study the effect of the cancer cells' competitive ability on the lifespan of the multicellular organism, see Fig.~\ref{fig4}(b).
In the absence of microenvironmental feedback, the lifespan of the multicellular organism decreases as $K$ increases.
It is intuitive that the higher the competitive advantage of cancer cells, the faster the fraction of them increases in the multicellular organism.
However, we obtain counterintuitive results with the presence of microenvironmental feedback.
We find that increasing $K$ extends the lifespan $L$ with microenvironmental feedback when cancer cell's competitive ability increases in $K\in \left\lfloor10, 20\right\rfloor$, see Fig.~\ref{fig4}(b).
The reason for this phenomenon is that functional cells can dynamically adjust their competitive ability according to the state of the microenvironment and the competitiveness of cancer cells, as illuminated in Eq.~\ref{1}.
To clearly explain this phenomenon, we show the evolution of the fraction of functional cells and the microenvironmental state for two choices of $K=10$ and $20$ in Fig.~\ref{fig5}.
We find that the relative magnitude of the reduction rate of the functional cells for $K=10$ and $K=20$ changes on both sides of $t=1795$.
The rate of decline of functional cells for $K=20$ is larger than that for $K=10$ when $t<1795$ in Fig.~\ref{fig5}(a).
During this period, the competitive ability of the cancer cells dominates the impact on functional cells.
And when $t>1795$, TME emerges faster for $K=10$, which leads to a decrease in the competitive ability of functional cells, and thus reduces the growth rate of the functional cells as shown in Fig.~\ref{fig5}(b). Larger $K$ gives those cancer cells more evident competitive advantage during the microenvironmental feedback, despite the slow emergence of TME.
Therefore, there exists an optimal lifespan for the organism by adjusting the competitive ability of cancer cells.

\subsection{The Parondo's paradox of cancer cells' competitive ability}
\label{The Parondo Paradox of cancer cells' competitive ability}

Recently, some counterintuitive phenomena are observed when organisms with different life history traits (strategies) compete with each other~\cite{Capp2021, Wen2022PRL}, which has been successfully explained by means of a game-theoretic framework known as the Parondo's paradox~\cite{Harmer1999, Harmer2000, Harmer2001}.
Basically, the Parondo's paradox describes an interesting, yet counterintuitive at a first glance, situation that the periodic and/or random switching of two losing strategies can lead to a successful one~\cite{Harmer1999, Harmer2000, Harmer2001}.

In Sec.~\ref{Effect of microenvironmental feedback parameters on lifespan.}, we have seen that both too low and too high competitive ability of cancer cells are losing strategies for the lifespan of multicellular organisms, since there is an optimal selection of $K$ for the multicellular organism to live the longest.
Then, similar to what has been done in~\cite{Wen2022PRL}, we wonder if a periodic/random switching between too low and too high competitive ability of the cancer cells could result in a greater lifespan than in each case alone, i.e., the emergence of the Parondo's paradox?

\begin{figure}[tb]
	\begin{center}
		\includegraphics[width=0.4\textwidth]{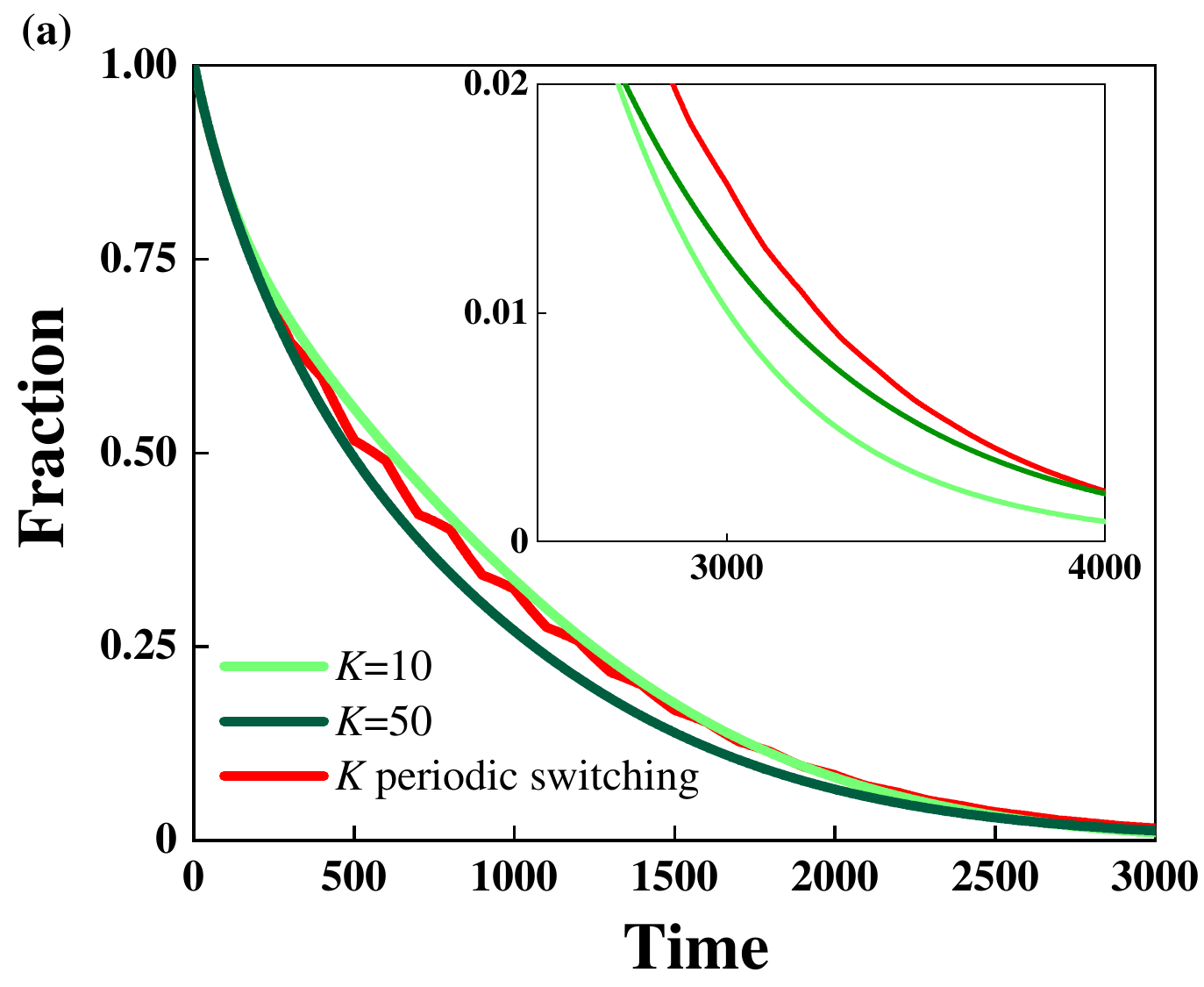}
		\includegraphics[width=0.4\textwidth]{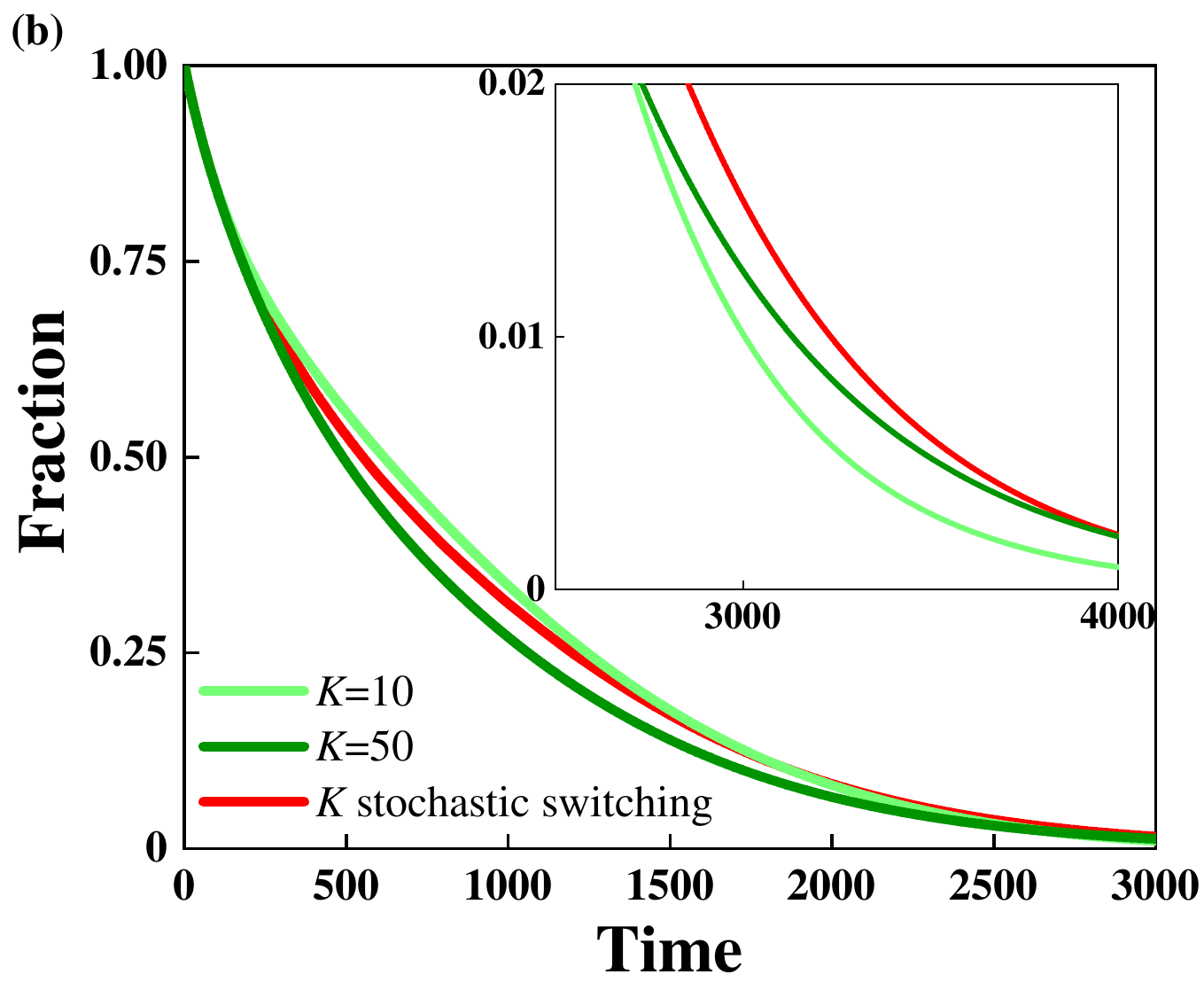}
		\includegraphics[width=0.4\textwidth]{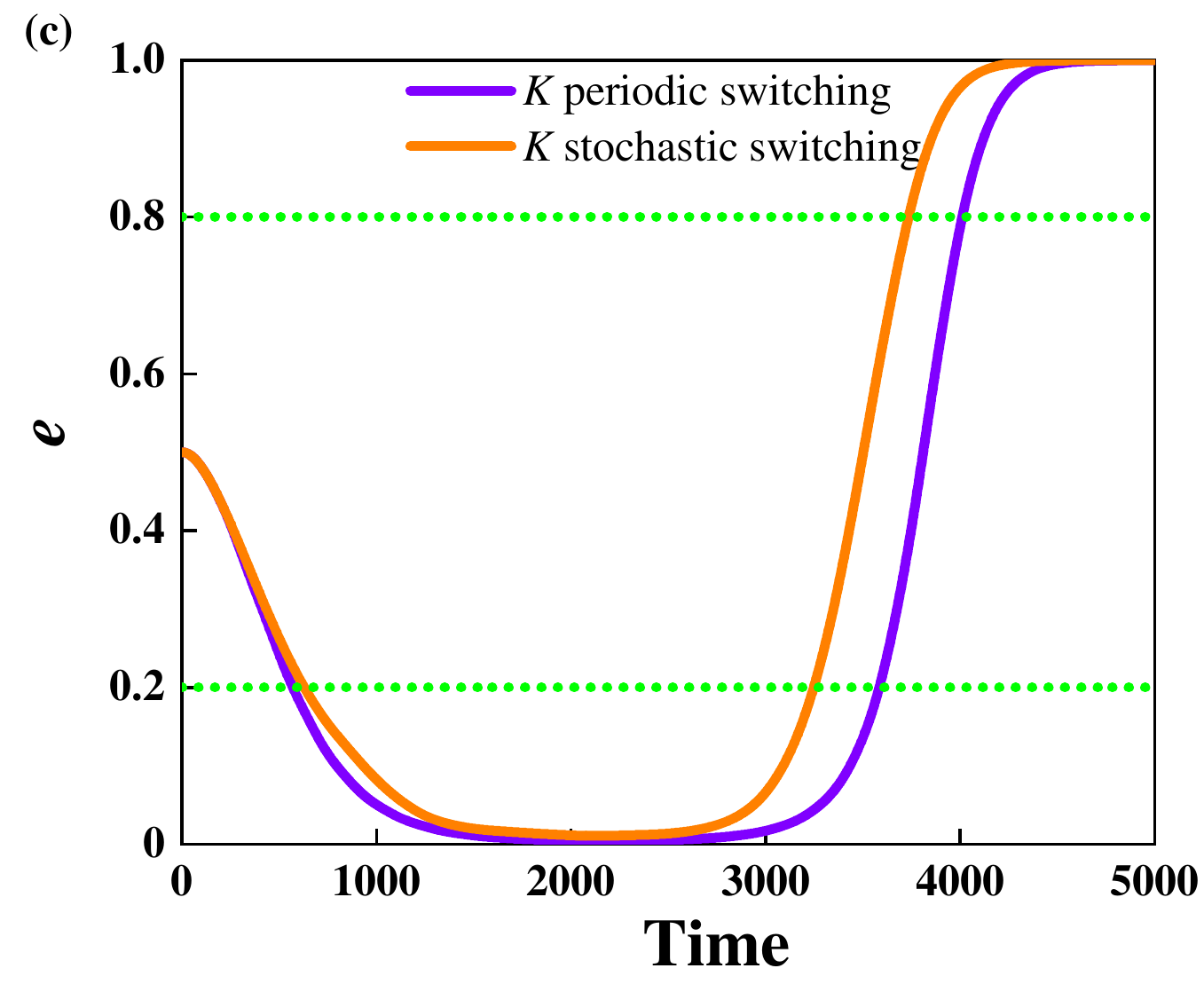}
		\caption{
Fraction of the functional cells under different switching schemes, where (a) is for the periodic switching with $\xi=100$ and $T=200$ ($L_{\mathrm{Periodic}}=3243$); and (b) is for the stochastic switching with $\tau=200$ and $\zeta=50\%$ ($L_{\mathrm{Stochastic}}=3234$).
(c) Time evolution of the microenvironmental state under different switching schemes.
Other parameters: $p=0.005$, $\theta=2$.}
		\label{fig6}
	\end{center}
\end{figure}

To answer this question, we first consider the case that the competitive ability of the cancer cells evolves as:
\begin{eqnarray}
	K & = & \left\{\begin{array}{ll}
		10, & 0 \leq  \text{mod} (t, T)<\xi, \\
		50, & \xi \leq  \text{mod} (t, T)<T.
	\end{array}\right.
\end{eqnarray}
Here, the competitive ability of the cancer cells switches periodically, which is $K=10$ during $[0,\xi)$ and $K=50$ during $[\xi,T)$.
Compared to the case of $K=10$ or $K=50$ alone, we find that a periodical switch of the competitive ability of the cancer cells can lead to a slower reduction of functional cells to $1\%$ as shown in Fig.~\ref{fig6}(a).
This reveals clearly that with the microenvironmental feedback mechanism, a strategy of periodic switching in the competitive ability of the cancer cells can extend the lifespan of multicellular organisms.

As an alternative to the periodic switching scheme~\cite{Wen2022PRL}, we also check whether a similar phenomenon can occur when the competitive ability of the cancer cells switches randomly?
To do this, we implement a stochastic switching scheme of the competitive ability of the cancer cells defined as
\begin{eqnarray}
	K  =  \left\{\begin{array}{ll}
		10, &   \text{mod} (t, \tau )=0, \; \text{prob}: \zeta, \\
		50, &   \text{mod} (t, \tau )=0, \; \text{prob}: 1-\zeta.
	\end{array}\right.
\end{eqnarray}
That is, the competitive ability of the cancer cells switches as $10$ or $50$ every $\tau$ time units, and the probability of choosing $10$ is $\zeta$.
The decrease of the competitive ability of the cancer cells may be caused, for instance, by radiotherapy and chemotherapy.
Figure~\ref{fig6}(b) displays the fraction of the functional cells under the stochastic switching scheme. Interestingly, the reduction in the fraction of the functional cells to $1\%$ is relatively longer than that in the case of $K=10$ and $K=50$ separately.
By doing some measurements, we have verified that the periodic switching scheme is more effective in promoting the lifespan $L$ than the stochastic switching, i.e., $L_{\mathrm{Periodic}}> L_{\mathrm{Stochastic}}$.
This is due to the faster speed of the appearance of TME in the case of stochastic switching of $K$, as shown in Fig.~\ref{fig6}(c).

Here, we want to emphasis that the presence of the microenvironmental feedback is crucial in giving rise to the interesting phenomenon of Parondo's paradox.
We have checked that without the microenvironmental feedback mechanism, the Parondo's paradox is absent for both the switching scheme of the competitive ability of the cancer cells, please refer to results in the Appendix~\ref{Switching of competitive ability between cancer cells without microenvironmental feedback}.

On the one hand, in the framework of intercellular competition with microenvironmental feedback, too low competitive ability of the cancer cells is a losing strategy because it cannot reverse the tendency of senescent cells to accumulate.
On the other hand, too high competitive ability for the cancer cells is also a losing strategy, because too high competitive ability makes it difficult for even functional cells to survive leading to the cancerous transformation of the organisms.
The arise of the Parondo's paradox suggest that the multicellular organisms (or new medicine treatment) may combine these two strategies to achieve the longer lifespan in the presence of pthe microenvironmental feedback.

\section{Conclusions and discussions} \label{conclusions}

In summary, we have studied the dynamical behavior of a mathematical model of intercellular competition, where a microenvironmental feedback mechanism is introduced to regulate the competitive ability of the different types of cells.
In the absence of microenvironmental feedback, our model produces results consistent with those found in Ref.~\cite{Nelson2017}, i.e., intercellular competition can weed out senescent cells, but the unlimited proliferation of cancer cells leads to inevitable aging.
With the involvement of the microenvironmental feedback, however, we find that the lifespan of the multicellular organisms can be effectively extended as long as the regulation rate of cancer cells on the microenvironment is not so strong.

It is well known that malignant tumors can greatly shorten the lifespan of the organisms~\cite{Campisi2009}.
In recent years, the average lifespan shortened by cancer is decreasing owning to the improvement of medical treatment.
Interestingly, the reduction in the lifespan revealed by our model matches quite well with the data for non-Hodgkin lymphoma in Canada from 1980 to 2015~\cite{Pham2020}, which suggests that our proposed model might serves as a good candidate to treat the interactions of different types of cells in real organisms.
Our results highlight specially the important role of the feedback rate of the fraction of cancer cells relative to the fraction of senescent cells to the microenvironment in extending the lifespan of the organism.
In the presence of the microenvironmental feedback, the development of the TME can be efficiently postponed.
In this sense, we contend that $\theta$ (i.e., the feedback strength of the cancer cells regulating the microenvironment, which could be characterized by the frequency and intensity of medical treatments in realistic situations) should be considered as a relevant index for measuring the effectiveness of treatments targeting the inhibition of TME.

Moreover, by exploring the effect of cancer cells' competitive ability on the lifespan of organisms, we have discovered a non-monotonic phenomenon caused by the microenvironmental feedback.
There is an optimal value of $K$ (the competitive ability of cancer cells) that allows the multicellular organisms to achieve the longest lifespan, which is totally absent without the microenvironmental feedback mechanism.
Of particular interesting, we have found that either periodic or stochastic switching between too high and too low values of $K$p can further extend the lifespan of the organism, as compared to the cases of too low or too high $K$ separately, which resembles like the phenomenon of Parrondo's paradox~\cite{Wen2022PRL,Harmer1999}.

Besides competitive interactions, complex cooperation among cells of multicellular organisms can further complicate the regulation of microenvironmental feedback~\cite{Wolf2018}.
Our current model describes just a homogenized microenvironment at the organismal level, which might be less of significance for applications in realistic situations.
In fact, it is very unlikely to reverse the microenvironment of the organism when it had already been dominated by the TME or the aging microenvironment~\cite{Arneth2020}.
Nevertheless, with the rapid development of modern medicine, we may expect at some future date that the microenvironment in real organism can be properly regulated in the procedure treating malignant tumors to achieve most effective treatment.
We thus hope that our theoretical work can inspire further experimental studies (such as \emph{in vitro} experimental systems~\cite{Fiorini2020}) to better describe the relationship between the microenvironment and intercellular competition, providing valuable clues on tumor therapy.

\section*{Acknowledgements}
This work is supported by the National Natural Science Foundation of China (Grants No. 11975111 and No. 12047501).
We thanks Bin-Quan Li and Cong Liu for useful discussions.

\appendix
\section{Impact of $\lambda$ and initial state of the microenvironment on the lifespan.}
\label{Impact of lambda and initial value of microenvironment state on lifespan.}

\begin{figure}[tb]
	\begin{center}
		\includegraphics[width=0.4\textwidth]{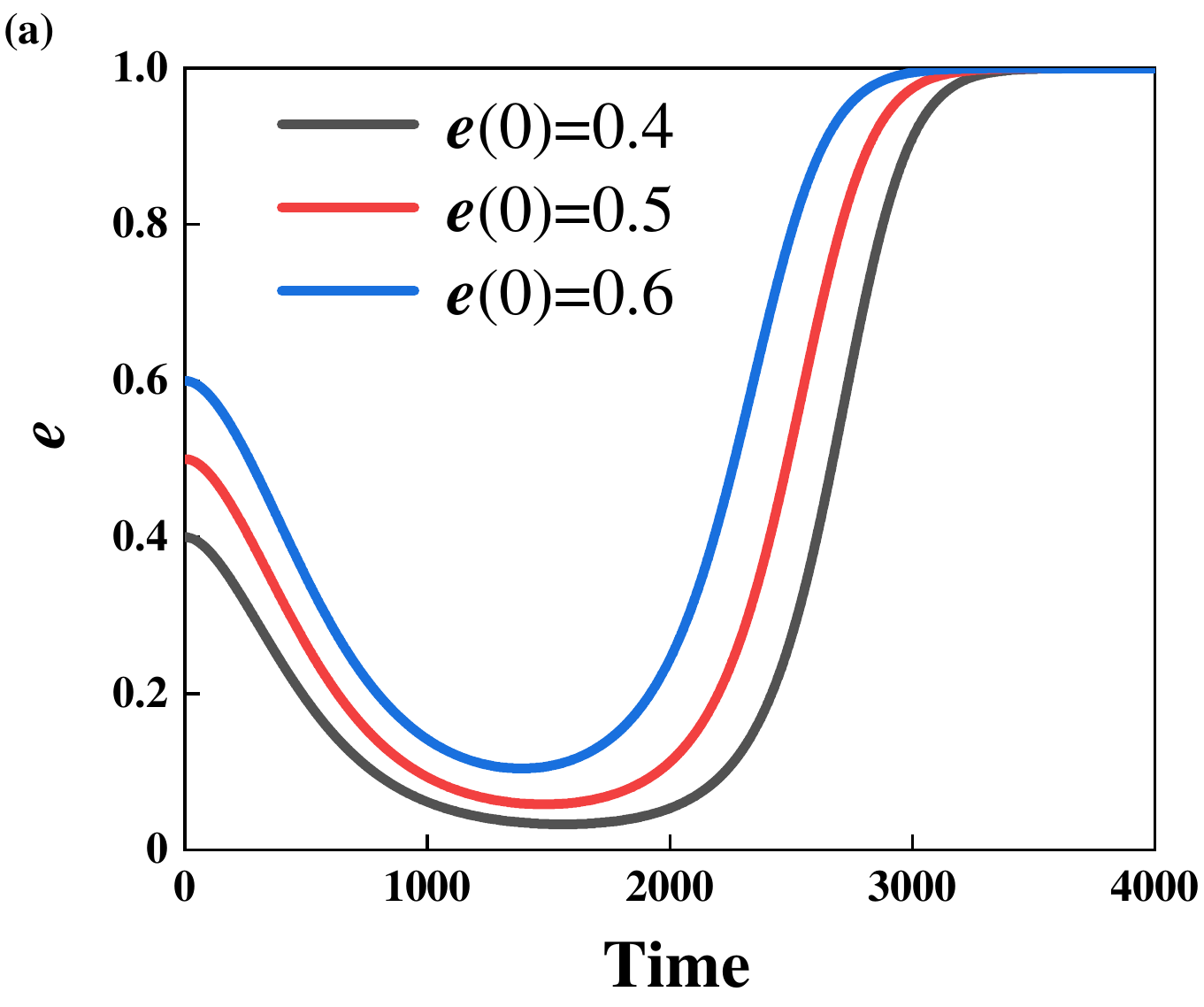}
		\includegraphics[width=0.4\textwidth]{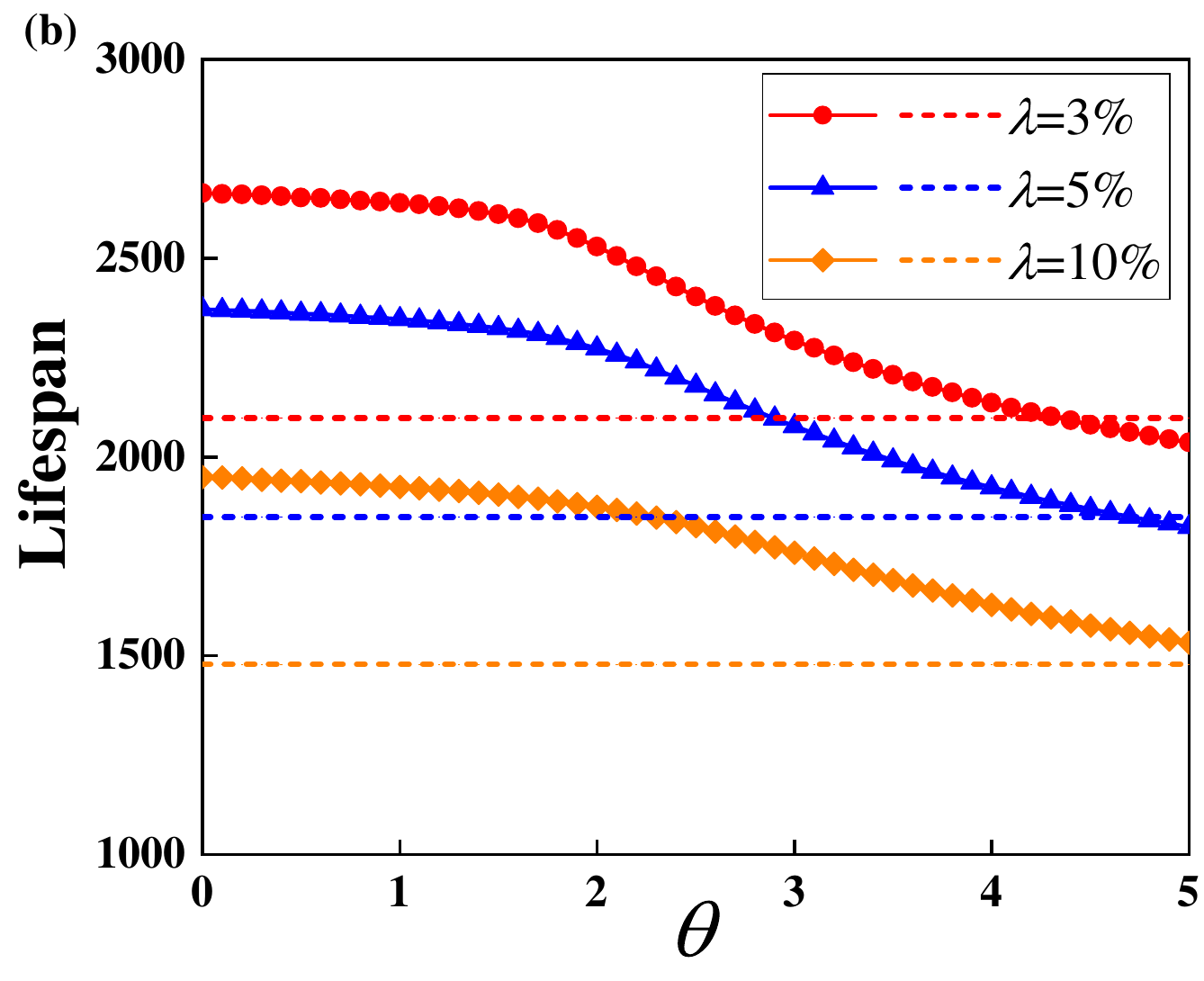}
		\caption{
(a) Time evolution of the microenvironmental state $e(t)$ starting from different initial values $e{(0)}$. Parameters: $p=0.005$, $\theta=2$.
(b) The lifespan $L$ as a function of the parameter $\theta$ for three choices of $\lambda = 3\%, 5\%, 10\%$.
The dotted lines mark the lifespan in the case of no microenvironmental feedback.}
		\label{fig7}
	\end{center}
\end{figure}

Accounting for the fact that organisms would be in microenvironmental homeostasis without aging and cancer development, we also investigate the effect of different initial states of the microenvironment [$e(0) = 0.4, 0.5, 0.6$] on its dynamical evolution in Fig.~\ref{fig7}(a).
We find that as the initial value increases, the microenvironmental state evolves more rapidly to the TME. Nonetheless, the evolutionary trend remains qualitatively the same, and for this reason we mainly focus on the results for $e(0)=0.5$ in the main text.

To verify whether the threshold phenomenon (i.e., for sufficiently large $\theta$, the lifespan $L$ fall blow the case where microenvironmental feedback is not taking into account) is robust to the other definition of the lifespan, we have calculated the lifespan $L$ as a function of $\theta$ for different choices of $\lambda=\{3\%, 5\%, 10\%\}$.
As $\lambda$ increases [Fig.~\ref{fig7}(b)], the observed threshold phenomenon always appears.
This indicates that the threshold behavior of $L$ versus $\theta$ originates from the microenvironmental feedback mechanism, and is a robust phenomenon.

\section{Switching of $K$ without microenvironmental feedback}
\label{Switching of competitive ability between cancer cells without microenvironmental feedback}

\begin{figure}[tb]
	\begin{center}
		\includegraphics[width=0.4\textwidth]{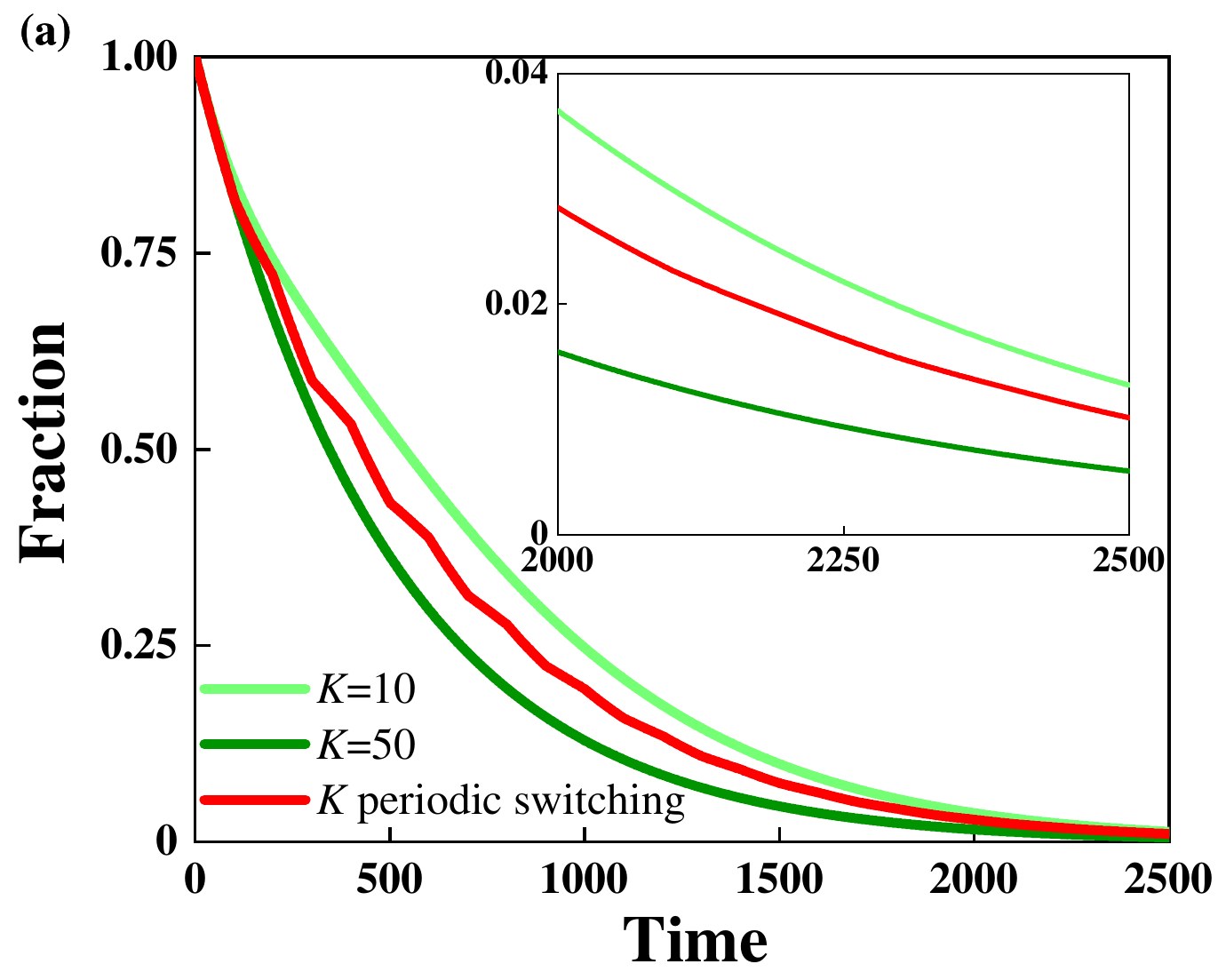}
		\includegraphics[width=0.4\textwidth]{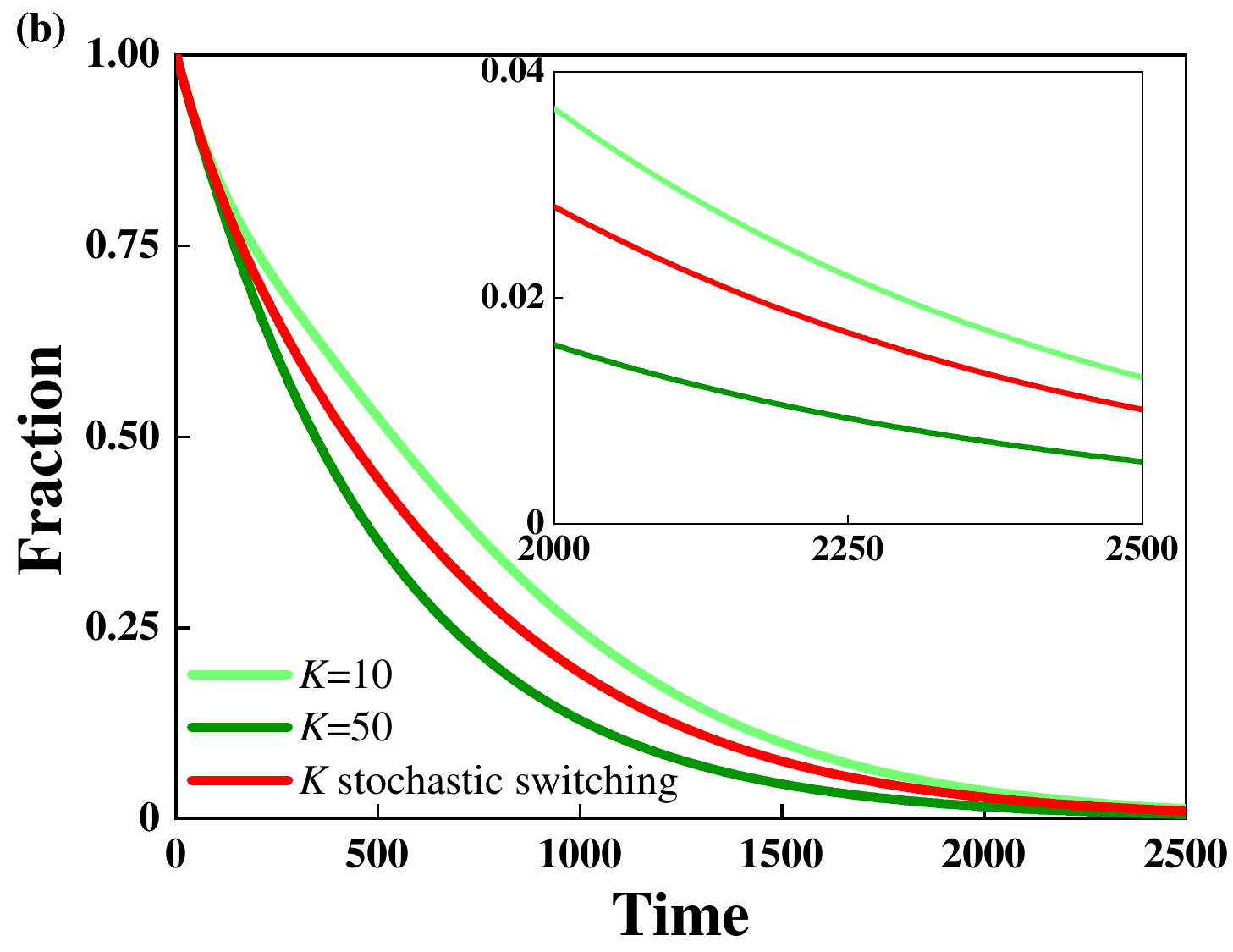}
		\caption{Fraction of functional cells under different switching schemes without microenvironmental feedback. (a) Periodic switching with $\xi=100$ and $T=200$. (b) Stochastic switching with $\tau=200$ and $\zeta=50\%$. Parameters: $p=0.005$.}
		\label{fig8}
	\end{center}
\end{figure}

We here complement the results when the competitive ability $K$ of the cancer cells switches between $10$ and $50$ in the absence of microenvironmental feedback ($\alpha_{F}\equiv 5.5$, $e(0) \equiv 0.5 $).
Figure~\ref{fig8} shows the fraction of the functional cells under periodic and stochastic switching.
The parameters of the two switching schemes are consistent with those in Fig.~\ref{fig6}.
However, we find that without microenvironmental feedback, the lifespan is just monotonically shortened when the competitive ability of the cancer cells increases.
This means that $K=10$ is a failed strategy, $K=50$ is a even more failed strategy, and there is no optimal situation in between.
Specifically, either periodic or stochastic switching of $K$ cannot lead to the arise of Parondo's paradox in the absence of microenvironmental feedback mechanism.

\bibliography{ref}

\end{document}